\documentclass[twocolumn,english,showpacs,superscriptaddress,prl]{revtex4-1}
\usepackage[latin9]{inputenc}
\usepackage{color}
\usepackage{float}
\usepackage{textcomp}
\usepackage{amstext}
\usepackage{graphicx}

\makeatletter

\providecommand{\tabularnewline}{\\}

\providecommand{\tabularnewline}{\\}
\usepackage{graphics}\usepackage{subfigure}\usepackage{longtable}\usepackage{pstricks}\usepackage{dcolumn}\usepackage{bm}

\makeatother

\usepackage{babel}
\begin{document}
\title{Bulk single crystal growth of the theoretically predicted magnetic
Weyl semimetals $R$AlGe ($R$ = Pr, Ce)}
\author{Pascal Puphal}
\email{pascal.puphal@psi.ch}

\affiliation{Laboratory for Multiscale Materials Experiments, Paul Scherrer Institute,
5232 Villigen, Switzerland}
\author{Charles Mielke}
\affiliation{Laboratory for Multiscale Materials Experiments, Paul Scherrer Institute,
5232 Villigen, Switzerland}
\author{Neeraj Kumar}
\affiliation{Paul Scherrer Institute, 5232 Villigen, Switzerland}
\author{Y. Soh}
\affiliation{Paul Scherrer Institute, 5232 Villigen, Switzerland}
\author{Tian Shang}
\affiliation{Laboratory for Multiscale Materials Experiments, Paul Scherrer Institute,
5232 Villigen, Switzerland}
\affiliation{Ecole Polytechnique F\'ed\'erale de Lausanne (EPFL), Lausanne CH-1015,
Switzerland. }
\author{Marisa Medarde}
\affiliation{Laboratory for Multiscale Materials Experiments, Paul Scherrer Institute,
5232 Villigen, Switzerland}
\author{Jonathan S. White}
\email{jonathan.white@psi.ch}

\affiliation{Laboratory for Neutron Scattering and Imaging, Paul Scherrer Institute,
5232 Villigen, Switzerland}
\author{Ekaterina Pomjakushina}
\email{ekaterina.pomjakushina@psi.ch}

\affiliation{Laboratory for Multiscale Materials Experiments, Paul Scherrer Institute,
5232 Villigen, Switzerland}
\begin{abstract}
We explore two methods for single crystal growth of the theoretically
proposed magnetic Weyl semimetals $R$AlGe ($R$ = Pr,Ce), which prove
that a floating zone technique, being both crucible- and flux-free,
is crucial to obtain perfectly stoichiometric $R$AlGe crystals. In
contrast, the crystals grown by a flux growth technique tend to be
Al-rich. We further present both structural and elemental analysis,
along with bulk magnetization and electrical resistivity data on the
crystals prepared by the floating zone technique. Both systems with
the intended 1:1:1 stoichiometry crystallize in the anticipated polar
\textit{I}4$_{1}$md (No. 109) space group, although neither displays
the theoretically expected ferromagnetic ground state. Instead PrAlGe
displays a spin-glass-like transition below 16\,K with an easy-c-axis
and CeAlGe has an easy-ab-plane antiferromagnetic order below 5\,K.
The grown crystals provide an ideal platform for microscopic studies
of the magnetic field-tunable correlation physics involving magnetism
and topological Weyl nodes. 
\end{abstract}
\maketitle

\section{Introduction}

In 1929, Hermann Weyl suggested a solution to the Dirac equation for
fermions in terms of massless particles, the so-called Weyl fermions\citep{Weyl(1929)}.
Only recently, however, has experimental evidence for their existence
as quasiparticles in (Ta,Nb)As \citep{Xu(2015),Lv(2015),Xu2(2015)}
been presented, leading to these compounds being known as Weyl semimetals.
The hallmark feature of these materials are topologically nontrivial
band touching points -- the Weyl nodes -- in their electronic spectra
\cite{Hasan(2015),Hasan(2017)}, which can endow the host system with
technologically promising properties in fields ranging from catalysis
\cite{Rajamathi2017,Politano2018} to optoelectronics \cite{Chan2017,Sie2019}.

The most versatile possibilities for Weyl node formation is in systems
displaying simultaneously broken spatial inversion (SI) and time reversal
(TR) symmetries. With TR symmetry broken naturally by the onset of
magnetic order, magnetic semimetals are ideal candidates for studying
the magnetic field-tunable correlation between magnetism and Weyl
physics {[}10{]}. Indeed, the influence of Weyl nodes on the transport
properties has been evidenced in some magnetic systems \cite{Suz(2016),Liu(2018),Kim(2018)},
motivating the study of the associated phenomena that is of both fundamental
and technological importance \cite{Tok(2017)}. 

From recent first principles theoretical calculations, it has been
predicted that the members of the $R$AlGe ($R=$ Pr, Ce) system are
new magnetic Weyl semimetals \cite{Chang(2018)} that offer remarkable
tunability, since the number and location of Weyl nodes may be controlled
by choice of the rare earth element \citep{Chang(2018)} and the types
of the broken symmetry, i.e. SI and/or TR, via the Al/Ge content.
In addition, in the presence of the combined broken symmetries, the
system offers a rich phase diagram that may be explored via self-doping
or chemical substitution. Therefore, to enable a broad range of experimental
studies on this class of material, there is a clear interest for establishing
the details for the growth of sizable ($\sim$mm$^{3}$) single crystals
and their basic physical characterization.

First discovered in 1992, $R$AlGe was initially described to crystallize
in the so called $\alpha-$ThSi$_{2}$ structure-type with a centrosymmetric
space group $I$4$_{1}$/amd (No. 141) \citep{Dhar(1992)}. Later
on however, it has been realized instead that $R$AlGe crystallizes
in the LaPtSi-type structure \citep{Dhar(1996),Gladys(2000)}, with
a body-centered tetragonal Bravais lattice and a polar, i.e. SI breaking,
space group $I$4$_{1}$md (No. 109). A subsequent study of the silicon
variants $R$Al$_{x}$Si$_{2-x}$ revealed a tunability of the structure-type
according to the Si content~\citep{Bobev(2005)}. It was established
that single crystals including all Lanthanides could be prepared in
quartz ampoules using the high-temperature flux technique with molten
Al as a solvent, and the flux removed by centrifugation \cite{Bobev(2005)}.
Due to both the evaporation and reaction of Al with quartz, a fast-cooling
rate was implemented in order to avoid a breaking of the ampoules.

To date, little is know about the physical properties of $R$AlGe.
For PrAlGe, there is only a detailed study of the crystal structure
\cite{Gladys(2000)}, with no physical property characterization.
For CeAlGe contradictory results are published for the magnetic properties;
early bulk susceptibility data show the magnetic Ce ions to order
below $\sim$6 K, but the system has been reported to order as a ferromagnet
(FM) \citep{Flandorfer(1998)} and an antiferromagnet (AFM) \citep{Dhar(1996)}.
Most recently, a study on flux grown CeAlGe single crystals was reported
\cite{Hodovanets(2018)}, with bulk magnetization measurements evidencing
ferromagnetic coupling in one direction, and antiferromagnetic coupling
in another, apparently resolving contradictions so far. In more detail,
the identified easy-axis of {[}100{]} suggested antiferromagnetic
order in the ab-plane.

Importantly however, the crystals studied in Ref. \cite{Hodovanets(2018)}
were prepared following the same route as reported in Ref. \cite{Bobev(2005)},
using SiO2 ampoules. As discussed in Ref. \cite{Bobev(2005),Hodovanets(2018)},
this choice of ampoule often leads to Si inclusions and a stable co-existing
CeAlSi phase, while at the same time, samples that are even slightly
Al-rich can lead to different crystal structures that maintains SI
symmetry \cite{Dhar(1996),Hodovanets(2018)}, thus greatly limiting
the propensity for the formation of topological Weyl nodes. Therefore,
a stoichioemtric analysis is necessary to ensure the intended 1:1:1
stoichiometry of $R$AlGe is achieved, and we will show that this
is less likely to be achieved in flux-grown samples compared with
crystals grown by the floating zone method. 

Here we present the successful single crystal growth of both PrAlGe
and CeAlGe, first by Al self-flux growth without the use of quartz
ampules, and secondly using crucible-free bulk crystal growth in a
floating zone in a mirror furnace. Only by the latter approach truly
stochiometric single crystals were obtained reliably. In addition,
we performed differential thermal analysis (DTA) to optimize the flux
profile and find the melting points. Finally we present both a magnetic
characterization and transport data obtained from the single crystals,
with the results confirming that the crystals have physical properties
consistent with those expected for magnetic semimetals.

\section{Experimental Details}

For the flux growth a home built tubular furnace was used that was
connected to vacuum and argon lines.

Thermogravimetric analysis was performed using a NETZSCH STA 449C
analyser.

The polycrystalline rods for the floating zone growth were cast in
a SCIDRE KTS - levitation melting facility. By induction melting,
the three starting elements Ce/Pr, Al and Ge of a minimum purity of
99.99\% are levitated in a strongly changing magnetic field, followed
by a sudden switching-off so that the melt falls into a cooled copper
shaper. The floating zone growth was performed in a SCIDRE HKZ - high
pressure, high-temperature, optical floating zone furnace.

The powder x-ray diffraction was performed using a Bruker D8 Advance
with a Cu cathode.

Energy dispersive X-ray spectra (EDS) were recorded with an AMETEK
EDAX Quanta 400{} detector in a Zeiss DSM 940A scanning electron
microscope (SEM).

Magnetic susceptibility measurements were carried out in a range of
1.8 - 400\,K and 0 - 7\,T using a Quantum Design Magnetic Property
Measurements System (MPMS). Resistivity measurements were carried
out for the range of 1.8 - 300\,K and 0 - 4\,T, as well as and AC
measurements from 10 - 20\,K on a Quantum Design Physical Property
Measurement System (PPMS).

\section{Crystal Growth}

\subsection{DTA analysis}

\begin{figure}[H]
\begin{centering}
\includegraphics[width=1\columnwidth]{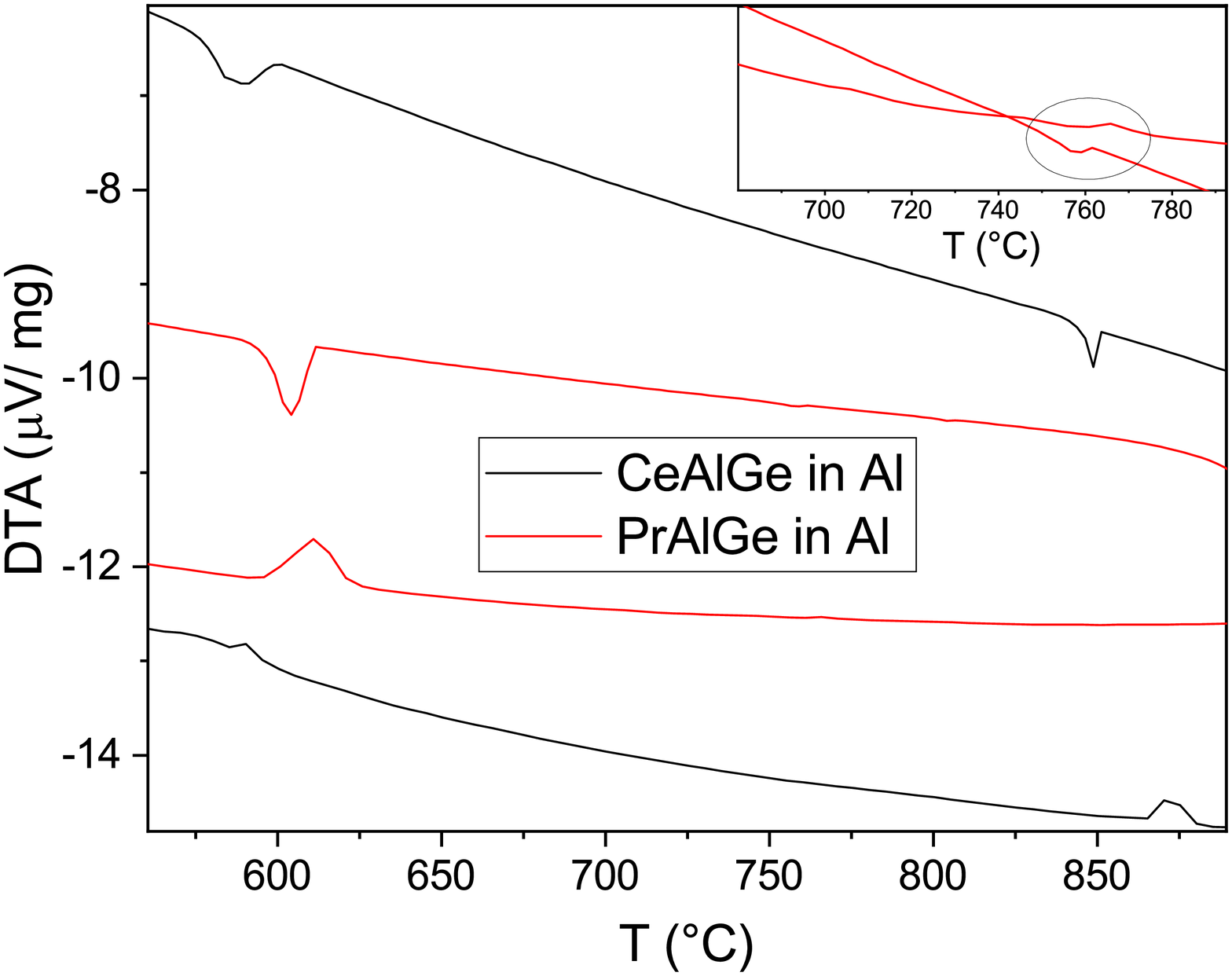} 
\par\end{centering}
\caption{\textcolor{black}{\label{DTA} Extract of DTA curves of polycrystalline
CeAlGe (black) and PrAlGe (red) samples measured in the presence of
20 moles of Al flux at a quick heating and cooling rate of 10\,K/min
in the range of 30�C to 950�C. The first signal around 600�C corresponds
to} the melting and crystallization of the Al flux. The inset shows
a magnification of the PrAlGe DTA curve.}
\end{figure}
For both CeAlGe and PrAlGe we performed differential thermal analysis
(DTA) runs solely on the stoichiometric 1:1:1 mixture of reacted materials
obtained by arc melting (see sub-section D) as well as in Al-flux,
both in a reducing atmosphere with a He(95\%)-H$_{2}$(5\%) gas flow
of 60 cc/min. For a measurement on the polycrystalline $R$AlGe samples
without flux (data not shown) we could see no pronounced DTA peaks
denoting a melting up to 1500�C, proving them to have a higher melting
point. However, the dissolution and crystallization points of CeAlGe
and PrAlGe in aluminum flux were found to be at 860�C (see black curves
of Fig. \ref{DTA}) and 760�C (see red curves of Fig. \ref{DTA}),
respectively.

\subsection{Flux growth}

\begin{figure}[H]
 \begin{centering}
\includegraphics[width=1\columnwidth]{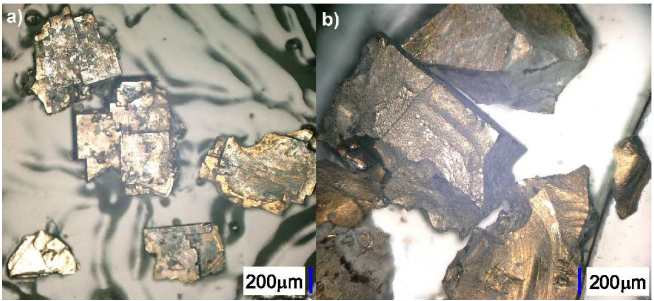} 
\par\end{centering}
\caption{\textcolor{black}{\label{flux} Pictures of the flux-grown crystals
of a) CeAlGe and b) PrAlGe right after flux removal using NaOH-H$_{2}$O,
and before subsequent annealing.}}
\end{figure}
Aiming to obtain sizeable $\sim$mm$^{3}$ crystals, we chose to perform
flux growth without quartz ampules, and used instead up-scaled alumina
crucibles of a large volume (100\,ml) and performed the growth under
argon atmosphere. This choice also provides the opportunity to refine
the previously published growth profile given in Ref. \cite{Bobev(2005)}.
There, the authors used a thermal ramping rate of 200�C/h to the reaction
temperature of 1175�C, held for 2\,h, followed by cooling to 700�C
at a rate of 30�C/h, where the Al flux was effectively removed by
centrifugation. In our approach, the flux was removed by dissolution
in a NaOH-H$_{2}$O solution.

Similarly to Ref.~\citep{Bobev(2005)}, we used a $R$:Ge:Al ratio
of 1:1:20 with 20\,g of aluminum granules, 5.2\,g Ce/ 5.3\,g Pr
pieces cut freshly from a rod in a He glove-box, and 2.69\,g Ge pieces.
These were all placed in a Al$_{2}$O$_{3}$ crucible that was placed
inside the tubular furnace. The optimal growth profile was found with
a quick heating of 300�C/h going up to 950�C, held for 2\,h and then
slowly cooled at 2.5�C/h to 685�C, followed by a quick cooling. We
chose a maximum temperature of 950�C since this higher temperature
compared to the dissolution point improved the homogenisation and
mixing. Afterwards, the whole piece displayed increased oxidation
on the surface due to both an oxide layer from the alumina granules,
and traces of oxygen in the argon gas due to the porosity of the furnace-tube,
but the crystals on the bottom were not affected. The crucible was
then placed in a NaOH-H$_{2}$O solution at 150�C until only the plate-like
crystals were left and could be filtrated. The general habit of the
crystals obtained was quite different between the two systems; for
CeAlGe (see Fig. \ref{flux} a) the crystals grow typically for tetragonal
systems in terrace shapes and as quite thin $c$-axis platelets, while
the PrAlGe crystals grow with sizable proportions along the $c$-axis,
and as bulky pieces (see Fig. \ref{flux} b). For both crystal species,
some aluminum flux on the surfaces remained that could not be removed
by the base, as further discussed in section IV. The crystals were
then put into fused silica ampules and annealed at 1100�C for one
week, leading to a darker appearance.

\subsection{Floating zone growth}

\begin{figure}[H]
 \begin{centering}
\includegraphics[width=1\columnwidth]{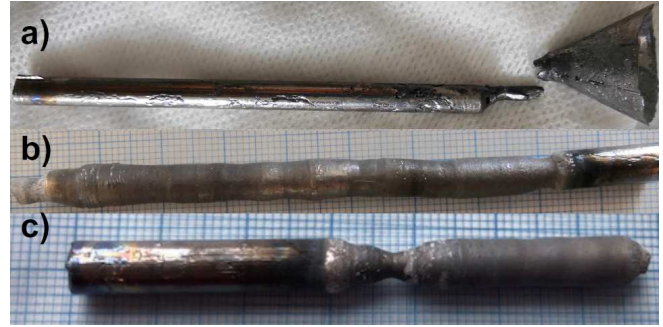} 
\par\end{centering}
\caption{\textcolor{black}{\label{float} Photos} of a) the cast CeAlGe rod,
and the floating zone grown crystals of b) CeAlGe and c) PrAlGe. }
\end{figure}
The starting rods for the floating zone crystal growth were cast using
a stoichiometric mixture of Ce or Pr, Al and Ge. Ce-pieces were freshly
cut from a rod in a He glovebox and corresponding amounts of Al and
Ge pieces were weighed. The materials were then transferred to the
levitation melting facility (KTS) equipped with a quartz tube. The
quartz chamber was flushed with Ar gas. 25\,g of starting material
was found to be sufficient for casting the rods using the KTS machine,
and rods were obtained like those shown in Fig.~\ref{float} a. CeAlGe
proves to have a higher melting point than PrAlGe; it starts to melt
at a power around 50\% of the 40\,kW and 100\,kHz generator with
a proper fluidity achieved at a power of 70\% enabling the casting.
In contrast, PrAlGe is homogeneously molten at just 45\% power. When
CeAlGe melts, a slight evaporation is observed, which is mainly due
to Al and is manifested as a tiny vapour pressure that develops at
1400�C, while Ce starts melting at 1900�C. The deduction that Al was
evaporating was confirmed by reduced Al content on repeatedly cast
rods measured by energy dispersive X-ray analysis (EDS).

The rods were then placed and centered in the high pressure, high-temperature
optical floating zone furnace (HKZ). After pumping the sapphire chamber,
a purification process of Ar was started in a flow of 0.1\,l/min.
The power was then slowly ramped up and we could pre-melt and connect
the rods at 30\% power of the 5kW Xenon lamp for CeAlGe and 20\% for
PrAlGe. A temperature check using a pyrometer lets us deduce a melting
point slightly above$\sim1500$�C for PrAlGe and $\sim1600$�C for
CeAlGe. For CeAlGe, we chose a higher pressure of 30\,bar argon compared
with 5\,bar for PrAlGe, to suppress the described evaporation. The
presence of $R$AlO$_{3}$ traces causes a particular behavior of
the melt whereby the oxide moves to the surface and forms a solid
layer around the liquid part, causing some shaking. However, it also
helped to keep a stable liquid, and enable a simple necking.

\begin{figure}[H]
 \begin{centering}
\includegraphics[width=1\columnwidth]{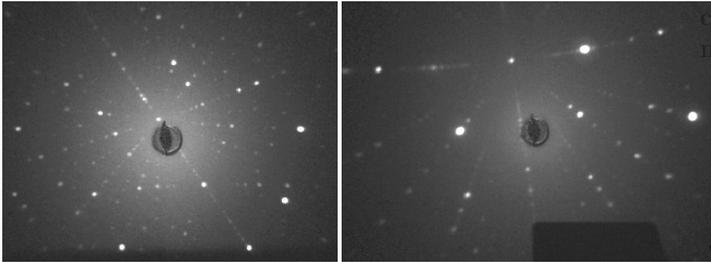} 
\par\end{centering}
\caption{\textcolor{black}{\label{Laue} Backscattered x-ray Laue images from}
the PrAlGe crystal taken along and perpendicular to the growth direction.
The images revealed a slight tilt of the a/b- (first image) and $c$-
(second) axis with respect to the principal growth axes.}
\end{figure}
By exploring different growth rates (1 - 20 mm/h), we have found that
in the case of PrAlGe a quick growth rate of 12 mm/h leads to two
dominant grains developing after a few millimeter of growth, which
grow equally quickly over the whole growth volume. For the case of
CeAlGe, such fast growth rates do not result in big crystalline grains.
We achieved the best result with a growth rate of 1\,mm/h, where
several (two or more) large mm sized grains develop over the very
last cm of the grown crystal from a total crystal length up to 10\,cm.

Due to the migration of $R$AlO$_{3}$ to the surface during the growth,
the resulting crystal appears less shiny than the starting polycrystalline
rod. This oxide layer can be removed easily by polishing. For both
materials, an x-ray Laue analysis revealed a tendency for growth along
an $a$-direction, with the c- and the second $a$-directions aligned
perpendicular to the rod surface. Being reasonably brittle, little
force is required to break the crystal, showing highly symmetric,
cleaved surfaces to separate the grains. We further note that CeAlGe
is less stable in air than PrAlGe, with a certain smell noticed when
being polished and sparks when being crushed. However, it oxidizes
slowly and builds a passivation layer similarly as pure Ce. Crystals
cut with a water-cooled diamond saw decayed partly into CeO$_{2}$
and broke into several pieces. PrAlGe, on the other hand, showed no
such features and seems to be more stable in air.

\subsection{Polycrystalline CeAl$_{x}$Ge$_{2-x}$}

To provide a better insight into the effect of stoichiometry, we produced
polycrystalline samples of CeAl$_{1.1}$Ge$_{0.9}$ and CeAl$_{0.9}$Ge$_{1.1}$
by arc melting the respective stoichiometric mixture of Ce, Al and
Ge pieces prepared in the same way as the rods. The samples were then
characterized by X-ray diffraction. Attempts to synthesize CeAl$_{1.5}$Ge$_{0.5}$
gave rise to CeAl$_{4}$ impurities, while the attempt to synthesize
CeAl$_{0.5}$Ge$_{1.5}$ yielded Ce$_{2}$Al$_{3}$Ge$_{4}$. We found
that phase pure samples could be obtained by reducing the substitution
level down to 10\% (CeAl$_{1.1}$Ge$_{0.9}$ and CeAl$_{0.9}$Ge$_{1.1}$).
As a last step, the stoichiometry was checked by EDS analysis to ensure
a homogeneous elemental distribution.

\section{Crystal structure}

\begin{table*}
\caption{\label{tab:Crystallographic-data-of}Crystallographic data of CeAlGe
and PrAlGe obtained by powder X-ray diffraction (PXRD) performed at
295\,K on crushed single crystals obtained both by the floating zone
and the flux-technique. Space group $I$4$_{1}$md (No. 109) with
all atoms in the (4a) position (0 0 z).}
\centering{}{\small{}}%
\begin{tabular}{c|cc|cc|cc|cc}
compound  & \multicolumn{2}{c|}{CeAlGe (floating zone)} & \multicolumn{2}{c|}{CeAlGe (flux)} & \multicolumn{2}{c|}{PrAlGe (floating zone)} & \multicolumn{2}{c}{PrAlGe (flux)}\tabularnewline
\hline 
EDS & \multicolumn{2}{c|}{Ce$_{1.02(7)}$Al$_{1.01(16)}$Ge$_{0.97(9)}$} & \multicolumn{2}{c|}{Ce$_{1.0(1)}$Al$_{1.12(1)}$Ge$_{0.88(1)}$} & \multicolumn{2}{c|}{Pr$_{1.08(24)}$Al$_{0.97(7)}$Ge$_{0.95(17)}$} & \multicolumn{2}{c}{Pr$_{1.0(1)}$Al$_{1.14(1)}$Ge$_{0.86(1)}$}\tabularnewline
a $\left[\textrm{�}\right]$  & \multicolumn{2}{c|}{4.28155(1)} & \multicolumn{2}{c|}{4.28978(4)} & \multicolumn{2}{c|}{4.25009(1)} & \multicolumn{2}{c}{4.26330(4)}\tabularnewline
c $\left[\textrm{�}\right]$  & \multicolumn{2}{c|}{14.6919(6)} & \multicolumn{2}{c|}{14.73222(16)} & \multicolumn{2}{c|}{14.62316(5)} & \multicolumn{2}{c}{14.69512(18)}\tabularnewline
$R$ factor  & \multicolumn{2}{c|}{4.59} & \multicolumn{2}{c|}{7.16} & \multicolumn{2}{c|}{2.99} & \multicolumn{2}{c}{7.59}\tabularnewline
 & z  & U  & z  & U  & z  & U & z  & U\tabularnewline
\hline 
$R$ & 0.59235  & 0.00533(2)  & 0.58991 & 0.0047(6) & 0.60832  & 0.0055(1) & 0.60725 & 0.00433(8)\tabularnewline
Al & 0.17736 & 0.01529(5)  & 0.17412 & 0.0248(14) & 0.19040 & 0.0212(3) & 0.19350 & 0.01014(18)\tabularnewline
Ge & 0.01049  & 0.01529(5)  & 0.00686 & 0.0248(14) & 0.02570 & 0.0212(3) & 0.02435 & 0.01014(18)\tabularnewline
\end{tabular}
\end{table*}
\begin{figure}[h]
 \begin{centering}
\includegraphics[width=1\columnwidth]{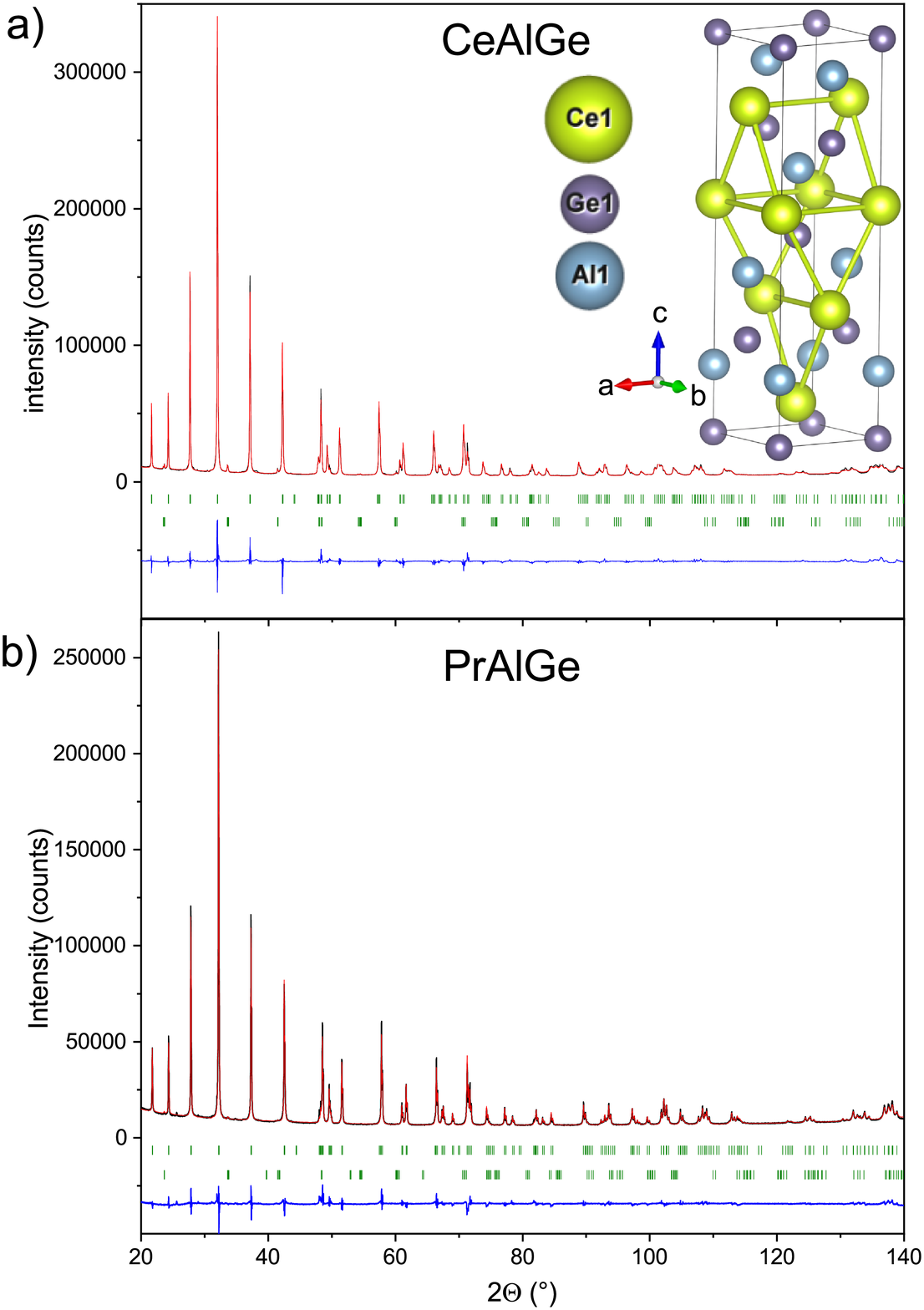} 
\par\end{centering}
\caption{\textcolor{black}{\label{XRD} }Rietveld refinement of the crystal
structure parameters of both compounds from PXRD data. The observed
intensity (black), calculated profile (red), and difference curve
(blue) are shown for crushed crystals obtained by the floating zone
technique of a) CeAlGe and b) PrAlGe. The rows of ticks at the bottom
correspond to the calculated diffraction peak positions of the phases
(from top to bottom): a) CeAlGe 97.7(5)\,wt\% and CeAlO$_{3}$ 2.3(1)\,wt\%
b) PrAlGe 99.3(4)\,wt\%, PrAlO$_{3}$ 0.7(0)\,wt\% . Inset: an image
of the refined $I$4$_{1}$md (No. 109) structure is shown.}
\end{figure}
As mentioned in the introduction, CeAlGe was first reported to crystallize
in the $\alpha-$ThSi$_{2}$ structure-type with space group $I$4$_{1}$/amd
(No. 141) \citep{Dhar(1992)}, with later studies instead proposing
the LaPtSi structure-type \citep{Dhar(1996),Gladys(2000)} with a
body-centered polar tetragonal space-group $I$4$_{1}$md (No. 109)
- see table \ref{tab:Crystallographic-data-of}. The latter structure
is shown in the inset of Fig. \ref{XRD} a). The difference between
this structure and the $\alpha-$ThSi$_{2}$ type is that while the
Al and Ge atoms occupy sites of different Wyckoff symmetry for the
$I$4$_{1}$md structure, they occupy a symmetry equivalent position
in the $I$4$_{1}$/amd case with 50\% occupation of each element
leading to an inversion center. Our refinement of PXRD data obtained
from stoichiometric crushed single crystals obtained by the floating
zone technique confirms the $I$4$_{1}$md (No. 109) spacegroup, as
it describes the data better than $I$4$_{1}$/amd. This is not only
borne out by the refinement with the $I$4$_{1}$md spacegroup yielding
a lower $R$-factor, but we also find that if one tries to include
site mixing in the $I$4$_{1}$md case, there is no Al on the Ge site
and vice versa. As expected, and as described in Section II C, a small
amount of oxidized $R$AlO$_{3}$ is detected. The larger impurity
phase fraction for CeAlGe compared to PrAlGe is due to the slow oxidation
of the compound in air, which is enhanced in powder samples owing
to the large surface. A refinement of the PXRD pattern on crushed
crystals obtained from Al-flux show a larger unit cell strongly suggesting
an enhanced Al content which has a larger crystal radius with 0.675\,$\textrm{�}$
compared to 0.67\,$\textrm{�}$ of Ge both in sixfold coordination.

\section{Elemental Analysis (EDS)}

Using a scanning electron microscope (SEM), we performed an EDS analysis
on several powder and single crystal samples prepared by all the above-described
methods. This characterization is crucial, since both the structure-type
and physical property transition temperatures can vary according to
small compositional variations of $R$Al$_{x}$Ge$_{2-x}$ \cite{Dhar(1996),Hodovanets(2018)}. 

\begin{figure}[t]
 \begin{centering}
\includegraphics[width=1\columnwidth]{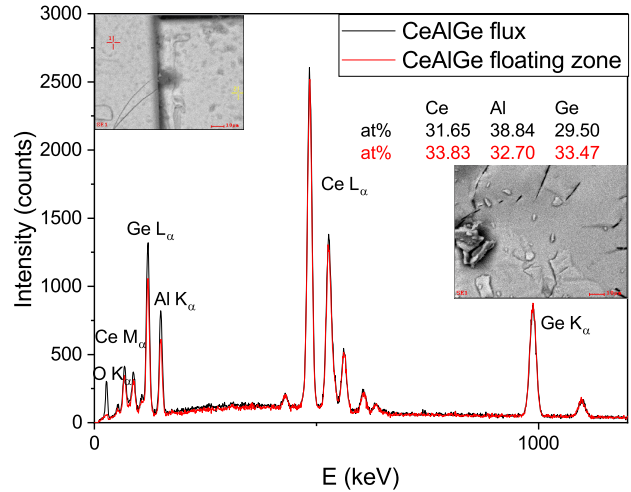} 
\par\end{centering}
\caption{\textcolor{black}{\label{EDX} Typical} EDS spectra from the surfaces
of two CeAlGe single crystals measured in an SEM with the corresponding
images as insets (topleft: flux grown, bottom right: floating zone
grown crystal).}
\end{figure}

As discussed in the introduction, flux growth has several issues,
e.g. if SiO$_{2}$ ampoules are used, Si can be incorporated into
the structure. We avoided this simply by performing the growth in
an Al$_{2}$O$_{3}$ crucible and under an argon flow. Due to the
application of Al self-flux, however, the tendency for substitution
of Ge by extra Al was observed in the crystals apparent from the EDS
and PXRD refinement results shown in table \ref{tab:Crystallographic-data-of}.
For CeAlGe grown by Al-flux the resulting average stoichiometry on
the surface was Ce$_{1.0(2)}$Al$_{1.3(5)}$Ge$_{0.7(3)}$ and on
a cleavage plane Ce$_{1.0(1)}$Al$_{1.12(1)}$Ge$_{0.88(1)}$. The
given values are averages taken over at least 5 points on several
samples, and the resulting statistical deviation is given in brackets.
In addition, we observe some remaining Al$_{2}$O$_{3}$ on the surface
of the crystals which can be seen as darker contrast in the SEM image
shown as the left inset of Fig. \ref{EDX}. Similar results are obtained
for PrAlGe samples with Pr$_{1.0(1)}$Al$_{1.2(2)}$Ge$_{0.8(2)}$
measured on the surface and Pr$_{1.0(1)}$Al$_{1.14(1)}$Ge$_{0.86(1)}$
 on a polished crystal. 

EDS measurements done on floating zone crystals shows them to be systematically much closer to the intended 1:1:1 stoichiometry, which can be expected for a crucible free growth in a congruently melting system, and where there is an absence of a mimimum evaporation of volatile compounds.
The average measured stoichiometries on cleaved surfaces as described in chapter III. C. are Ce$_{1.02(7)}$Al$_{1.01(16)}$Ge$_{0.97(9)}$ and Pr$_{1.08(24)}$Al$_{0.97(7)}$Ge$_{0.95(17)}$. The polycrystalline samples of 10\% Al over and underdoping show a measured stoichiometry of Ce$_{1.02(1)}$Al$_{1.14(1)}$Ge$_{0.84(1)}$ and Ce$_{1.06(5)}$Al$_{0.83(12)}$Ge$_{1.11(6)}$, respectively. 

\section{Bulk magnetic and transport Properties}

\subsection{CeAlGe}

While there is no publication on magnetic properties of PrAlGe, bulk
magnetic data from CeAlGe is mentioned in Refs.~\citep{Dhar(1992),Dhar(1996),Hodovanets(2018)}.
Due to the crystal field, Ce$^{3+}$ ions are expected to form a doublet
ground state that carries an effective spin-1/2, and which are coupled
via the Ruderman-Kittel-Kasuya-Yosida (RKKY) interaction. Previously,
an AFM transition was reported at $T_{N}=4$\,K \citep{Dhar(1992),Dhar(1996)}
or $\sim5$\,K \citep{Hodovanets(2018)}, with a relatively low entropy
of 5.3\,J/mol K \citep{Dhar(1992)} or 0.75\,R\,ln2 $\approx$
4.32\,J/mol K \citep{Hodovanets(2018)} compared with the theoretical
value of 5.76\,J/mol K for a spin 1/2 system. This effect should
not be associated with Kondo screening since the Sommerfeld coefficient
($\sim$20 mJ/mol K \citep{Dhar(1992),Dhar(1996)}) and resistivity
measurements indicate a normal magnetically ordered 4$f$ compound.
Therefore, we assign the missing entropy to lie within the low temperature
tail $\rightarrow$ 0\,K. The reported Curie-Weiss temperatures on
polycrystals are $\Theta_{W}=-13.5$\,K \citep{Dhar(1996)} or $\Theta_{W}=-3.6$\,K
\citep{Hodovanets(2018)} and an effective moment of $\mu_{eff}\approx2.57\thinspace\mu_{B}$
\citep{Dhar(1996),Hodovanets(2018)} indicates a Ce$^{3+}$ valence.

\begin{figure}[h]
 \begin{centering}
\includegraphics[width=1\columnwidth]{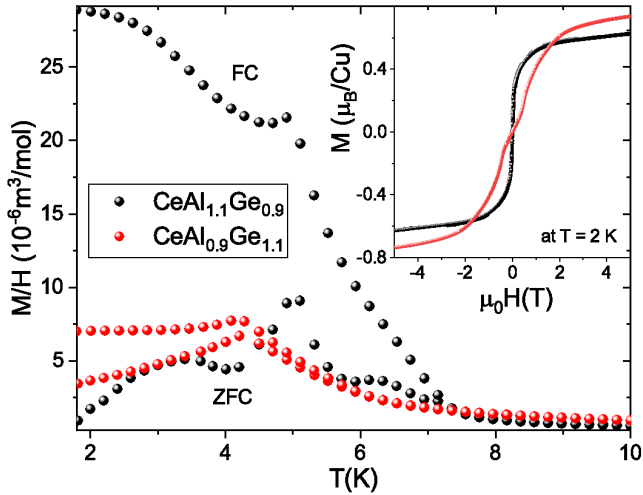} 
\par\end{centering}
\caption{\textcolor{black}{\label{Ce difference} Low temperature magnetization
measured in zero field cooled and field cooled manner on a powder
sample of 22\,mg ($x=1.1$) and 20\,mg ($x=0.9$) of CeAl$_{x}$Ge$_{2-x}$
in a field of 5\,mT. The inset shows the corresponding field dependence
of the magnetization measured at 2\,K.}}
\end{figure}
When discussing bulk measurements, we show first that the exact values
of characteristic physical properties are sensitive to the precise
stoichiometry, as was discussed previously, for example, in the context
of Fig. 4 in Ref. \citep{Dhar(1996)}. Therefore, the stoichiometric
variations between the samples provides a natural explanation for
differences in properties reported in the literature. To provide a
better insight into the effect of stoichiometric differences, we present
first magnetic measurements of polycrystalline samples of CeAl$_{1.1}$Ge$_{0.9}$
and CeAl$_{0.9}$Ge$_{1.1}$. The temperature-dependent magnetic susceptibility
of the samples was measured in the MPMS and the resulting low temperature
part is shown in Fig. \ref{Ce difference}. The data show the first
magnetic ordering transition temperatures indeed display a pronounced
dependence on the stoichiometry. For the Al-deficient sample, we observe
a kink in the susceptibility denoting an antiferromagnetic transition
near 4\,K, similarly as reported in Ref. \cite{Dhar(1996),Dhar(1992)}.
On the other hand, the Al-rich sample shows a more irregular thermal
behavior, showing a ferromagnetic-like transition near 7\,K, followed
by an antiferromagnetic-like one close to 5\,K. As found in field
dependent-magnetization a metamagnetic transition at 3.6\,T occurs
in the Al-deficient variant, which is absent in the Al-rich sample.
This strong sensitivity of the magnetic properties to the stoichiometry
clarifies the reason for a conflicting picture of the ground state
provided by previous reports, and proves the necessity for well-characterized
crystals when exploring physical phenomena in such samples.

With this in mind, we focus now on the characterization of relatively
large single crystals obtained via the floating zone technique.

\begin{figure}[h]
 \begin{centering}
\includegraphics[width=1\columnwidth]{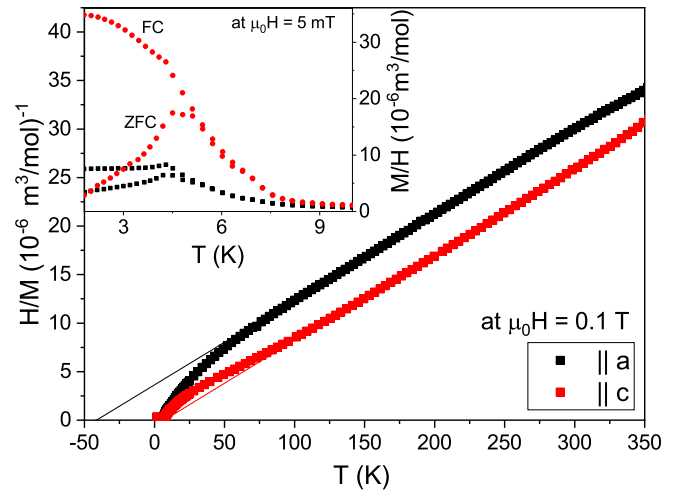} 
\par\end{centering}
\caption{\textcolor{black}{\label{susc Ce} Magnetic data obtained on a floating
zone grown CeAlGe single crystal with a mass of 125.4\,mg. }The magnetic
susceptibility was measured in the range of 1.8 - 400\,K with the
field aligned parallel to a (black dots) and parallel to c (red dots).
The main figure shows the inverse susceptibility obtained after field
cooling at 0.1 \,T. Inset: The low temperature range of both zero
field cooled and field cooled magnetic susceptibility curves measured
in 5\,mT. }
\end{figure}
In Fig. \ref{susc Ce} the temperature-dependent inverse susceptibility
H/M of CeAlGe measured in a field of 0.1\,T is shown for H along
both the a- and c-axes. Curie-Weiss fits of the high temperature part
of the data reveal a ferromagnetic Weiss-temperature along the $c$-axis
with $\Theta_{W}\approx10$\,K and an antiferromagnetic one along
the $a$-axis with $\Theta_{W}\approx-42$\,K. In both cases, the
effective moment obtained by fitting the high temperature range is
$\mu_{eff}\approx2.69\thinspace\mu_{B}$, this being slightly larger
than the theoretical one of $\mu_{eff}^{Ce^{3+}}=2.54~\mu_{B}$. In
the inset of Fig. \ref{susc Ce}, the low temperature part of the
susceptibility for both field directions in an applied field of 5\,mT
is shown after both field-cooling (FC) and zero-field-cooling (ZFC).
In each case, a maximum is observed indicating an AFM order to onset
at 4.5\,K (field parallel to c) and 4.3\,K (field parallel to a).
The magnetic anisotropy is also visible in terms of a stronger difference
between FC- and ZFC-curves for fields parallel to c.

\begin{figure}[h]
 \begin{centering}
\includegraphics[width=1\columnwidth]{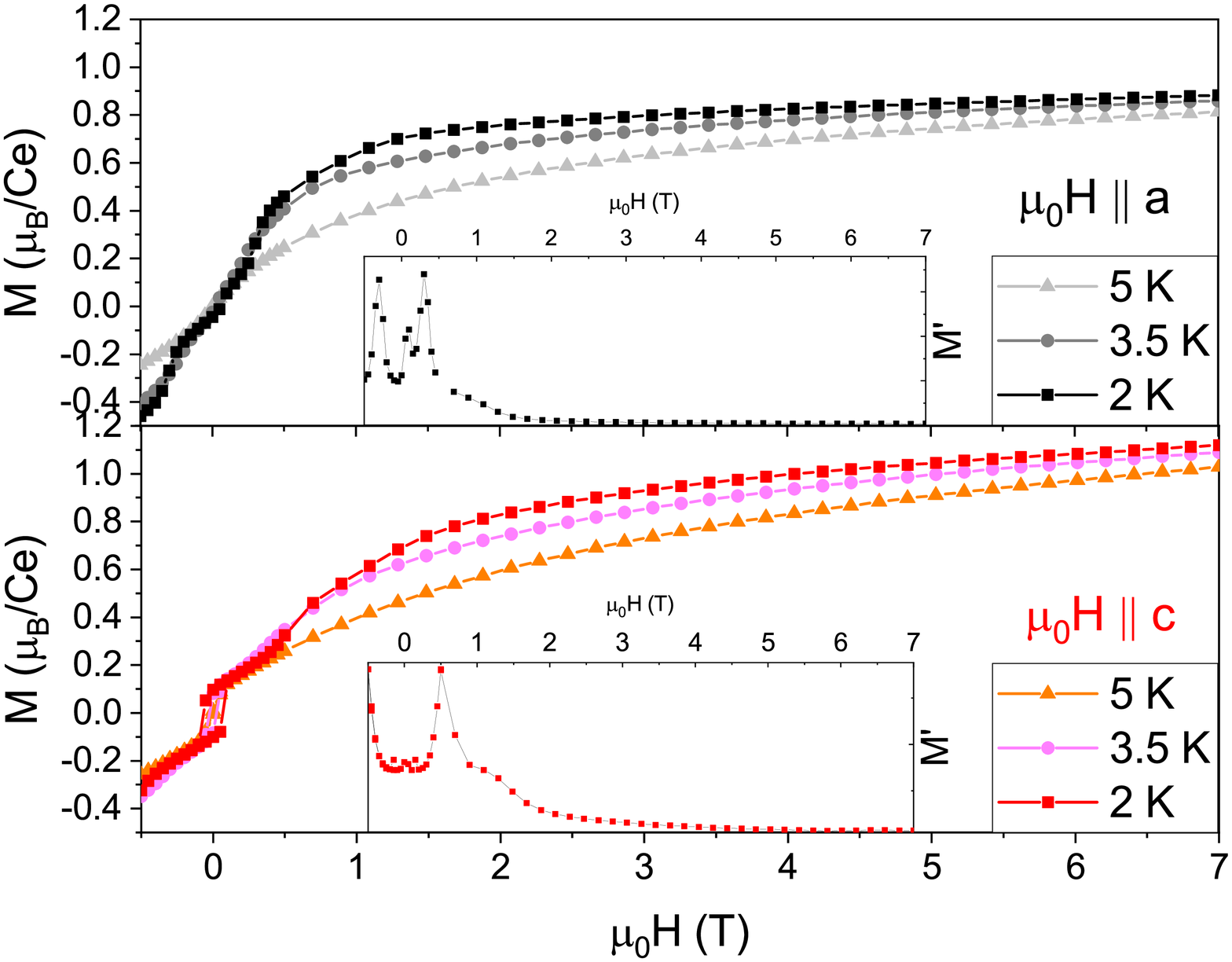} 
\par\end{centering}
\caption{\textcolor{black}{\label{susc Ce-1} Field-dependence of the magnetization
measured on a 125.4\,mg floating zone grown CeAlGe single crystal
at 2\,K, 3.5\,K, and 5\,K for both field directions: parallel to
a (black/gray symbols) and c (red/pink/orange symbols). Insets show}
the magnetic field derivative of the corresponding 2\,K dataset.}
\end{figure}
In Fig. \ref{susc Ce-1}, the field dependent magnetization is shown
for the field both parallel to a and c, and at three different temperatures
of 2\,K, 3.5\,K, and 5\,K around the AFM transition. For both field
directions, a metamagnetic transition is apparent, as seen clearly
in the field-derivatives shown as corresponding insets in Fig. \ref{susc Ce-1}.
The obtained values for the metamagnetic transition fields are $\sim$0.3\,T
for $H$ \textbar{}\textbar{} a and $\sim$0.6\,T for $H$ \textbar{}\textbar{}
c. A further transition for fields applied parallel to a is observed
at a lower field of around 0.1\,T, similarly as reported on the flux-grown
samples \citep{Hodovanets(2018)}. At 7~T, the saturation magnetization
$M_{s}$ is not reached as we still observe a slight increase; at
2~K the value of the magnetization is larger along the $c$-direction
with 1.12 $\mu_{B}$/Ce$^{3+}$ compared to 0.88 $\mu_{B}$/Ce$^{3+}$
for fields along the $a$-direction. A small hysteresis is visible
with a coercivity of $h_{c}\sim70$\,mT and a remanence of 0.11\,$\mu_{B}$
for fields along c.

\begin{figure}[h]
 \begin{centering}
\includegraphics[width=1\columnwidth]{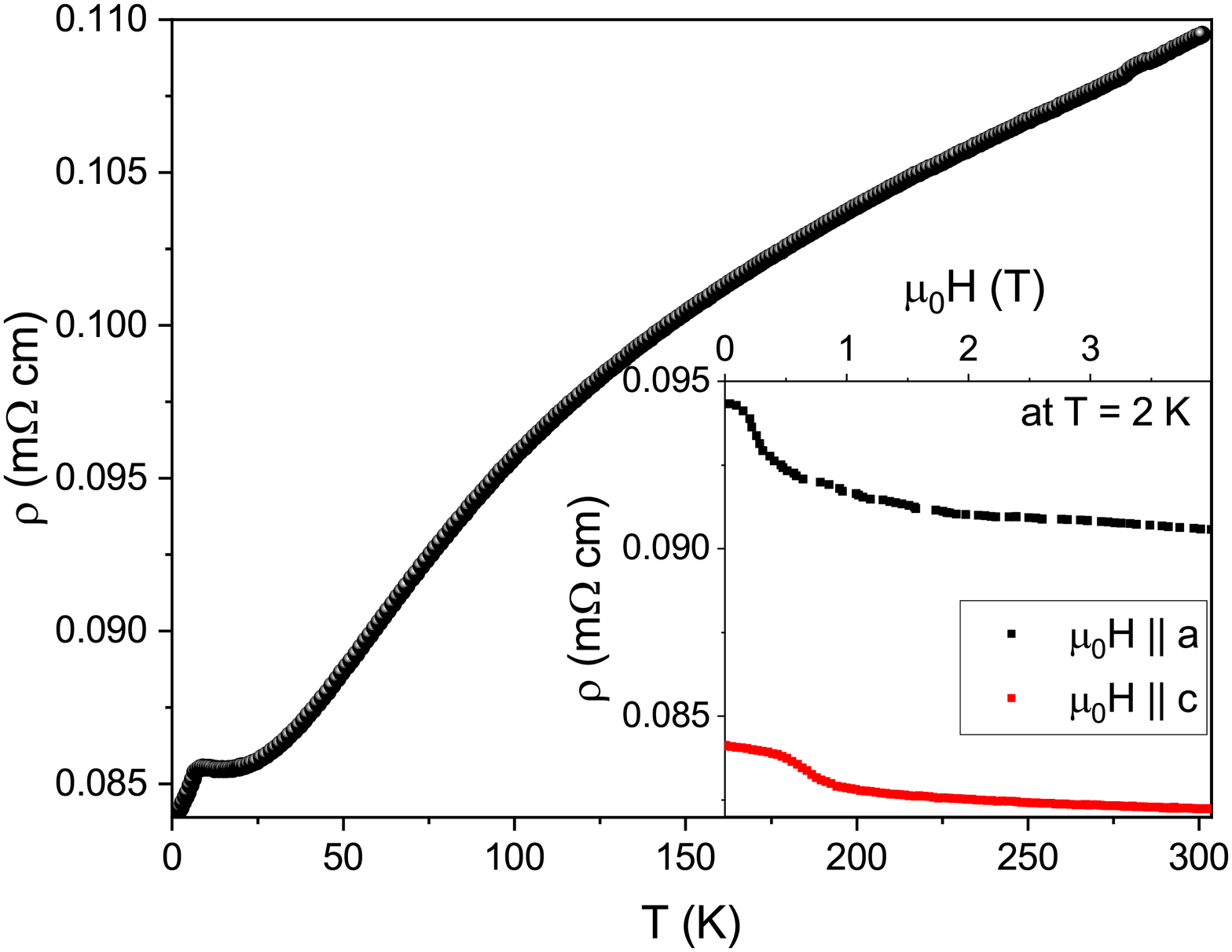} 
\par\end{centering}
\caption{\textcolor{black}{\label{res Ce} Temperature dependence of the resistivity
with the current running along the $a$-direction on a needle like
cut of 2.09\,mg from a floating zone grown CeAlGe single crystal.
The inset shows the field dependent resistivity with fields parallel
to c and the current along a (red symbols), and fields parallel to
a while the current runs along c (black symbols)}.}
\end{figure}
The temperature dependent restistivity data in zero field for a floating
zone grown CeAlGe crystal piece is shown in Fig. \ref{res Ce}. The
estimated residual resistivity ratio (RRR) is 1.3 which is a relatively
low value even for a semimetal \cite{Chen(2018)}, nonetheless it
is comparable to that reported for flux grown crystals of 2.3 \citep{Dhar(1996)}
and 2 \citep{Hodovanets(2018)}. The magnetic transition is visible
as a slight maximum in the expected temperature region. In the bottom
right inset of figure \ref{res Ce}, the field dependent resistivity
for fields applied both along the a- and the c-axes is shown to display
a negative magnetoresistance. These data show anomalies at the metamagnetic
transitions at around 0.3\,T (H \textbar{}\textbar{} a) and 0.6\,T
(H \textbar{}\textbar{} c) in agreement with measurements obtained
by $M(H)$.

\subsection{PrAlGe}

\begin{figure}[h]
 \begin{centering}
\includegraphics[width=1\columnwidth]{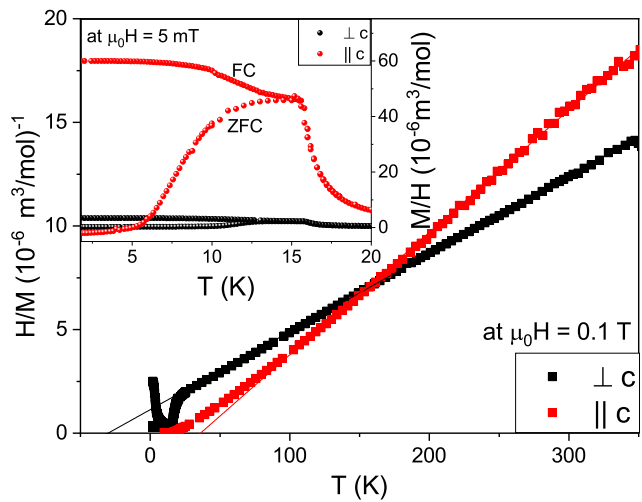} 
\par\end{centering}
\caption{\textcolor{black}{\label{susc Pr}Bulk magnetic property data obtained
on a PrAlGe single crystal. The} susceptibility was recorded in the
range of 1.8 - 400\,K with the field aligned perpendicular (black
dots) and parallel to c (red dots). The main panel shows the inverse
susceptibility measured after FC at 0.1 \,T. The inset depicts the
low temperature range of both ZFC and FC curves measured at 5\,mT.}
\end{figure}

In Fig. \ref{susc Pr}, we show the temperature dependent inverse
susceptibility H/M of a PrAlGe crystal measured either with a field
of 0.1\,T along the a- and $c$-axis. A Curie-Weiss fit of the data
reveals a ferromagnetic Weiss-temperature along the $c$-axis with
$\Theta_{W}\sim36$\,K and an antiferromagnetic one in the ab plane
with $\Theta_{W}\sim-30$\,K. The resulting effective moments are
$\mu_{eff}^{c}\approx4.1\thinspace\mu_{B}$ and $\mu_{eff}^{a}\approx3.3\thinspace\mu_{B}$
with a resulting powder value of $\mu_{eff}=2\mu_{eff}^{a}+\mu_{eff}^{c}=3.57\thinspace\mu_{B}$
close to the theoretical one of $\mu_{eff}^{\text{Pr}^{3+}}=3.58\thinspace\mu_{B}$.
The inset of Fig. \ref{susc Pr} shows the low temperature FC- and
ZFC-susceptibility obtained for a 5\,mT field applied along two crystallographical
directions. A sharp increase of the low temperature susceptibility
upon cooling is followed by a cusp-like transition at 16\,K. This
observation, along with a pronounced difference between FC- and ZFC-curves
for each field direction suggests the cusp to denote a spin-glass-like
transition. The data also reveal an Ising-like anisotropy with easy-axis
along the $c$-axis, and a second anomaly discerned around 11\,K
perhaps indicating a spin reorientation. The saturation magnetization
$M_{s}$ for fields along c is reached by $\sim$0.5\,T at 2.45 $\mu_{B}$/Pr$^{3+}$.
For fields in the ab plane, the magnetization increases slowly, reaching
a value of just 0.82\,$\mu_{B}$/Pr$^{3+}$ by 5\,T. For both field
directions, a hysteresis is observed with a coercivity of $h_{c}\sim60$\,mT/
$\sim0.1$\,T and a remanence of $\sim1.135$\,$\mu_{B}$ for fields
along c and $\sim$0.04\,$\mu_{B}$ for fields in the ab plane.

\begin{figure}[h]
 \begin{centering}
\includegraphics[width=1\columnwidth]{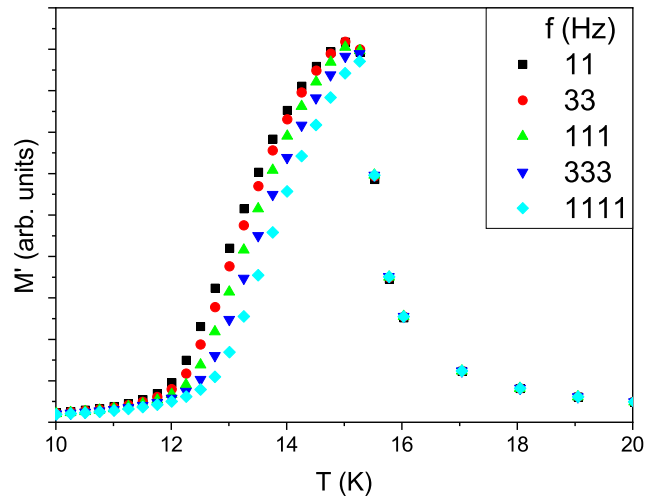} 
\par\end{centering}
\caption{\textcolor{black}{\label{AC} Temperature dependent AC magnetisation
measured on a PrAlGe single crystal at a frequency of 11, 33, 111,
333, 1111\,Hz with respect to the crystal oriented along the $a$-direction.}}
\end{figure}

Besides the indication for a spin-glass state below 16\,K given by
a large splitting of FC- and ZFC-magnetization, we observed a frequency-dependent
ac susceptibility of the cusp anomaly at 16\,K. The data in figure
\ref{AC} show the cusp anomaly moves to higher temperatures as the
ac frequency is increased, consistent with typical behavior for a
spin- or cluster-glass \cite{Mydosh2015}. Taken together, the magnetic
measurements suggest PrAlGe to enter a spin-glass state below 16\,K,
the origin of which may be due to frustration induced by competing
ferromagnetic and antiferromagnetic interactions. 

\begin{figure}[h]
 \begin{centering}
\includegraphics[width=1\columnwidth]{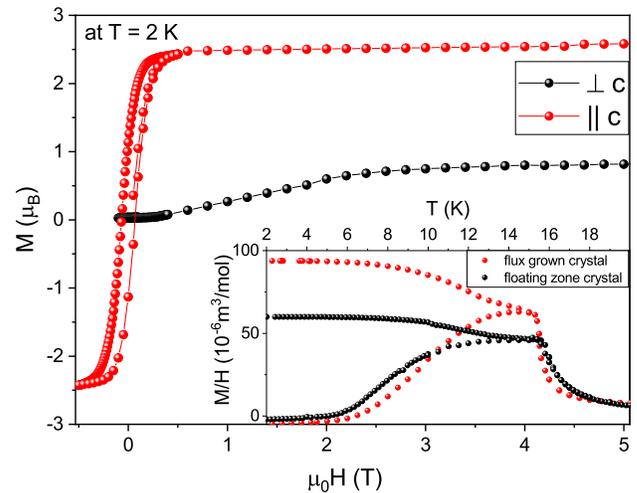} 
\par\end{centering}
\caption{\textcolor{black}{\label{susc Pr-1} Field dependent magnetisation
measured on a PrAlGe single crystal at 2\,K after ZFC, for both field
directions: perpendicular (black) and parallel to c (red). The inset
shows a comparison of the low temperatures susceptibility of a flux
grown crystal (5.4\,mg) with a floating zone grown crystal (101\,mg)
measured at 5\,mT with the field applied along the $c$ axis.}}
\end{figure}

Similarly as for CeAlGe, PrAlGe single crystals obtained via the flux
method proved to be Al-rich. However, the influence of stoichiometric
variance is not as strong as with the Ce ions since there is no structure-type
transition in the PrAl$_{x}$Ge$_{2-x}$ ($0.8<x<1.4$) series \cite{Gladys(2000)}.
The general magnetic properties of both flux and floating zone grown
PrAlGe crystals are similar, but a slight shift of the transition
temperature is apparent (see the inset of figure \ref{susc Pr-1}). 

\begin{figure}[h]
 \begin{centering}
\includegraphics[width=1\columnwidth]{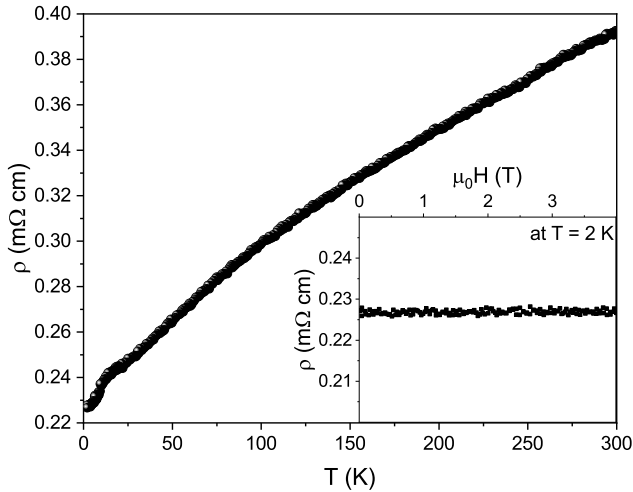} 
\par\end{centering}
\caption{\textcolor{black}{\label{res Pr} Temperature dependent resistivity
with a current along the $a$-axis of an oriented piece cut from a
77.3~mg floating zone grown PrAlGe single crystal. The inset shows
the field dependent resistivity with a field aligned along the $c$-axis
(H\textbar{}\textbar{}c).}}
\end{figure}

Temperature dependent resistivity data obtained for a floating zone
grown PrAlGe crystal of mass 77.3\,mg is shown in Fig. \ref{res Pr}.
The resulting residual resistivity ratio (RRR) is estimated to be
1.73, and thus similar to CeAlGe. The spin-glass-like transition is
visible in the resistivity as a slight kink just below $\sim16$\,K.
The inset of figure \ref{res Pr} shows the field dependent resistivity
for fields applied along the $c$-axis to be featureless over the
explored range measured at 2\,K.

\section{Summary}

To summarize, we have presented a methodology to grow large crystals
of the magnetic Weyl semimetal candidates $R$AlGe ($R$=Ce,Pr) using
the floating zone technique. Due to the inherent flexibility of $R$Al$_{x}$Ge$_{2-x}$,
it proves challenging to obtain a perfect 1:1:1 stoichiometry, and
slight variations influence both the structural as well as the physical
properties. Nevertheless, we reveal both CeAlGe and PrAlGe to crystallize
in the SI-breaking polar tetragonal $I$4$_{1}$md structure, thus
satisfying the expectation that these systems host Weyl fermions in
their magnetically-ordered ground states. The typical RRR values in
between 1 - 2 and a resistivity of $\sim0.1$\,m$\Omega$ cm suggest
the two materials to be semimetals.

From bulk magnetic characterization, we find PrAlGe to be spin-glass-like
below 16\,K with a spin reorientation around 11\,K and an easy-axis
along c. CeAlGe displays a rich magnetic phase diagram characterized
by an antiferromagnetic ground state below 5\,K, with the moments
lying in the ab-plane. We find neither of the systems to display the
simple ferromagnetic ground state expected according to theory \cite{Chang(2018)}.
This renders the anticipated topological phase more complicated, but
may nonetheless open the door to new phenomena in topological semimetals
linked to complex magnetic ground states. In this context the prepared
crystals provide the foundations for future detailed characterization
of the magnetic ground states by transport and microscopic probes
aimed at revealing the relation between magnetism and the properties
of the Weyl state. The availability of large crystals enables their
rich characterisation by microscopic probes such as neutron and photon
scattering. 
\begin{acknowledgments}
We thank C. Krellner for the provision of measurement time using the
SEM. D. Destraz and J. Chang for discussions and support from L. Berardo,
I. Zivkovic and H. R�nnow. The authors would like to acknowledge the
Swiss National Science Foundations (SNSF R\textquoteright Equip, Grant
No. 206021\_163997 and Grant No. 206021\_139082) and matching funds
from Paul Scherrer Institute for purchasing SCIDRE HKZ - high pressure
high-temperature optical floating zone furnace, the SCIDRE KTS - levitation
melting facility and the MPMS as well as funding from the SNSF Sinergia
network \textquotedblleft NanoSkyrmionics\textquotedblright{} (grant
no. CRSII5-171003).
\end{acknowledgments}

\bibliographystyle{apsrev4-1}
\addcontentsline{toc}{section}{\refname}\nocite{*}
\bibliography{weyllibrary}

\end{document}